\documentstyle[12pt]{article}
\begin{document}
\include{psfig}
\newcommand{\bq}{\begin{equation}}
\newcommand{\eq}{\end{equation}}
\newcommand{\bqa}{\begin{eqnarray}}
\newcommand{\eqa}{\end{eqnarray}}
\newcommand{\nl}{\nonumber\\}
\newcommand{\eqn}[1]{Eq.(\ref{#1})}
\newcommand{\skl}[1]{\vspace{#1\baselineskip}}
\newcommand{\apw}{a-priori weights}
\newcommand{\vx}{\vec{x}}
\newcommand{\gix}{g_i(\vx)}
\newcommand{\al}{\alpha}
\newcommand{\bal}{\bar{\al}}
\newcommand{\ai}{\al_i}
\newcommand{\gx}{g(\vx)}
\newcommand{\fx}{f(\vx)}
\newcommand{\sumn}{\sum_{i=1}^n}
\newcommand{\wx}{w(\vx)}
\newcommand{\aib}{\bal_i}
\newcommand{\bi}{\beta_i}
\newcommand{\oi}{\omega_i}
\newcommand{\thix}{\theta_i(\vx)}

\pagestyle{empty}
\begin{flushright}
NIKHEF-H 94-17\\
INLO-PUB-4/94
\skl{1}
\end{flushright}
\begin{center}
\begin{Large}
{\bf Weight optimization\\ in multichannel Monte Carlo\footnote{This
research has been partly supported by EU under contract number
CHRX-CT-0004.\\}}
\skl{1}
Ronald Kleiss\footnote{e-mail: t30@nikhefh.nikhef.nl}\\
NIKHEF-H, Amsterdam, the Netherlands\\
\skl{1}
Roberto Pittau\footnote{e-mail: pittau@rulgm0.leidenuniv.nl}\\
Instituut-Lorentz, University of Leiden, the Netherlands\\
\end{Large}
\skl{3}
{\bf Abstract}
\end{center}
We discuss the improvement in the accuracy of a Monte Carlo
integration that can be obtained by optimization of the
{\em a-priori weights\/} of the various channels. These
channels may be either the strata in a stratified-sampling
approach, or the several `approximate' distributions such as
are used in event generators for particle phenomenology.
The optimization algorithm does not require any initialization,
and each Monte Carlo integration point can be used in the evaluation
of the integral. We describe our experience with this method
in a realistic problem, where an effective increase in program
speed by almost an order of magnitude is observed.

\newpage
\pagestyle{plain}
\setcounter{page}{1}

In almost all Monte Carlo integrations, an effort must be made to
reduce the variance of the integrand \cite{hamhan,james}.
One of the currently popular approaches of variance reduction
is the so-called {\em stratified sampling\/} technique,
where the integration region is divided in a number of bins, with
a (usually) predetermined number of integration points in each
bin. An example of this technique is the program {\tt VEGAS} \cite{vegas},
in which the bins are automatically redefined from time to time
so as to reduce the integration error. Another approach is that of
{\em importance sampling\/}, where various techniques are used to
obtain (pseudo-)random variables that have a non-uniform rather than a
uniform distribution: one tries to generate a density of Monte Carlo points
that is larger in those parts of the integration region where also
the integrand is large, thus reducing the error (note that this
will work only when the integrand has large {\em positive\/} values;
large {\em negative\/} values do not lend themselves to probabilistic
modelling). Importance sampling is widely used in {\em event generators\/}
for particle phenomenology: also, the Metropolis algorithm \cite{metro}
used in statistical physics is actually a form of importance sampling.\\

In the construction of event generators for particle phenomenology, the aim
is usually to generate Monte Carlo points in some phase space of final-state
momenta (and spins), with a density proportional to a predetermined
multidifferential cross section. Often, such a cross section exhibits,
in different regions of phase space, peaks that find their best description
in terms of different sets of phase space variables.
An example is provided by brems\-strah\-lung in a particle collision process,
where the bremsstrahlung quanta are emitted, in the different Feynman
diagrams, by different particles.
It is customary, in such a case, to generate each peaking structure with
a different mapping of (pseudo-)random numbers: the particular mapping
used to generate an event is then chosen randomly, using a predtermined
set of probabilities, which we shall call {\em \apw \/}. It is the aim of
this paper to indicate how these \apw\ lend themselves to optimization.\\

First, we establish some notation. The function to be integrated is
$\fx$, where $\vx$ denotes a set of phase-space variables.
Each distinct mapping of random numbers into $\vx$ is called
a {\em channel\/}, and each channel gives rise to a different (non-uniform)
probability density, that we denote by $\gix$,
$i=1,2,\ldots,n$; $n$ is the number of channels in our {\em multichannel\/}
Monte Carlo. Each density $g_i$ is of course
nonegative and normalized to unity:
$\int\gix d\vx=1$. The \apw\ are denoted by $\ai$, and also these must be
a partition of unity: $\ai\ge0$ and $\sumn\ai=1$. If the channels
are picked at random, with probability $\ai$ for channel $i$, the total
probability density of the obtained sample of $\vx$ values is
$\gx=\sumn\ai\gix$, which is also
nonegative and normalized to unity.
Note that we may take the $\gix$ to be linearly independent.
The {\em weight\/}
assigned to each Monte Carlo point must, then, be $\wx=\fx/\gx$.
The expectation value of the result of this Monte Carlo integration,
and its variance, then follow from
\bqa
I & = & \langle\wx\rangle = \int d\vx\;\gx\;\wx = \int d\vx\;\fx\;\;,\nl
W(\al) & = & \langle\wx^2\rangle =
\int d\vx\;\gx\;\wx^2 = \int d\vx\;{\fx^2\over\gx}\;\;.
\eqa
Here we have indicated the dependence of $W$ on the set $\ai$.
The expected error of the integration, for $N$ points, is
$[(W(\al)-I^2)/N]^{1/2}$.
It is the quantity $W(\al)$ that we may try to minimize by adjustment
of the $\ai$. Note that, since $I$ does not depend on
$\gx$, we may change the $\ai$ during the integration, even from one Monte
Carlo point to the next. Therefore, any such optimization procedure will
always lead to an unbiased result.

The extremum of $W(\al)$ (homogeneous of degree -1 in the $\ai$)
on the simplex described by the $\ai$ is obtained for those values
$\aib$ for which $W_i(\bal)=W(\bal)$ for all $i$, where
\bq
W_i(\al) \equiv -{\partial\over\partial\ai}W(\al)
= \int d\vx\;\gix\;\wx^2\;\;.
\eq
That this extremum is, indeed, a minimum can be proven simply. For
let $\ai=\aib+\bi$, where $\bi$ is small. Then, $\sumn\bi=0$, and we have
\bq
W(\al) = W(\bal)+{1\over2}\int d\vx\;
{\fx^2\over\gx^3}\;\left(\sumn\bi\gix\right)^2
+{\cal O}(\bi^3)\;\;.
\eq
In general, we cannot prove that the minimum is unique.
It is interesting to study some simple cases. In the first place,
suppose that $\fx$ is dependent with the set $\gix$, that is, there
are constants $\gamma_i\ge0$ such that $\fx=\sumn\gamma_i\gix$.
We then have $\int d\vx\fx=\sumn\gamma_i$, and the minimum
for $W(\al)$ is reached at $\bal_k=\gamma_k/\sumn\gamma_i$, in which
case $W_k(\al)=W(\al)=(\sumn\gamma_i)^2$. Of course, in this case the
Monte Carlo error is zero.

A second case of interest is that of stratified sampling. This is
described, in our formalism, by a set of channels $\gix$ that each
restrict the values $\vx$ to a piece of phase space, a {\em bin\/}.
The bins must be disjoint and together make up the whole phase space volume.
That is,
\bq
\gix = {1\over\oi}\thix\;\;,
\eq
where $\thix$ is the characteristic function of bin $i$, and $\oi$ its
volume, $\oi=\int d\vx\thix$. In that case,
\bq
W_i(\al) = {\oi\over\ai^2}\int d\vx\;\thix\;\fx^2\;\;,
\eq
and we recover the well-known result \cite{james} that the error
is minimized if each bin contributes the same amount to the total
variance. An illustrative limiting case is that where all the bins are
vanishingly small and have equal volume $\omega$, so that
$\int d\vx\fx\thix=\omega f(\vx_i)$, where $\vx_i$ is the center of each
bin. We then have, again, that the optimal case is $\gx\propto\fx$.

A remark is in order here. Stratified sampling, as usually defined,
uses a deterministic, rather than a random, choice of the channels.
Typically, one first generates points in bin 1, then in bin 2, and so on.
In our formalism, this is covered by going over, for the variable
that determines the choice of channel, from a pseudorandom
variable to a strictly uniform one. Not surprisingly, one then has to
{\em predetermine\/} the exact number of Monte Carlo points, since
a uniform point set can only be generated in a deterministic
way if the total number of points is known. Whether this is an
attractive or a repulsive feature of our formalism, is largely
a matter of taste.\\

The following procedure suggests itself. Start the Monte Carlo, using
{\em some\/} set $\ai$, picked either randomly or on the basis of some
information on the behaviour of $\fx$ and $\gx$. After generating
a number of Monte Carlo points (a few hundred, say) in each channel,
estimate the $W_i(\al)$, using
\bq
W_i(\al) = \left\langle{\gix\over\gx}\wx^2\right\rangle\;\;.
\label{wiest}
\eq
Then, use this information to find an improved
set of $\ai$. Repeat this procedure until no further improvement is found;
the results of all iterations may be added in the final
integral estimate.
Obviously, when some $W_i$ is large (small), the optimization should
give a new $\ai$ that is larger (smaller) than the old one.
The above examples naturally suggest an optimization
prescription. In the case of stratified sampling, the optimum
was reached when $W_i(\al)=c\ai^{-2}$,
with some constant $c$ independent of $i$. This implies that,
supposing the idealized case where each $W_i(\al)$ is estimated with
zero error (this does {\em not\/} mean that the integral has zero error!),
the choice of new $\ai$ according to
\bq
\ai^{\mbox{{\small new}}}\;\;\propto\;\;\ai\sqrt{W_i(\al)}\;\;,
\eq
will give immediately the optimal set $\aib$, irrespective of the
choice of the initial set $\ai$. Therefore, we propose to use this
optimization prescription. It should be observed that the
extra computational burden is actually quite small, since,
whichever the actual channel is picked to generate an $\vx$,
one always has to compute the contributions from all channels
in the calculation of $\gx$.

There are a few points to be noted here. In the first place, the
new $\ai$ have to be renormalized to as to sum to unity: hence, it
is irrelevant whether we use $W_i$, or $W_i/W$, or
$W_i/\sum_{k=1}^n W_k$, or any other scaled version. Secondly,
in practice the values of the $W_i$ will have their own Monte Carlo
error, and the convergence will be less than immediate, even for
stratified sampling. This may actually be considered an advantage, since
numerical methods with slower convergence tend, in many cases,
to be the more robust ones.
Thirdly, for non-stratified sampling, the $\ai$
will influence each other, and our prescription cannot be shown to
be the best possible. However, in practice, the different
$\gix$ usually put their largest mass in
quite different regions of phase space. Channels that overlap to
a large extent will show a lot of `cross-talk', but then again,
the way they split their a-priori weights between them is less important
for the overall error (note that if a set of channels is linearly
dependent, only the sum of their \apw\ will be determined).\\

As in all numerical optimization schemes, many alternatives and
improvements of approach can be envisaged. We mention only two.
In the first place, one may decide to use, in the determination of $W_i$,
only the points actually generated with the corresponding channel $\gix$,
or using the points from all channels. The last choice,
which we prefer, and to which Eq.(\ref{wiest}) corresponds, has the advantage
that the error on $W_i$ will be smaller, and the drawback that then
these errors will be correlated. In the second place, nothing forces
one to actually {\em use\/} a putative updated set of \apw: on the basis
of a given sample of Monte Carlo points $\vx$, together
with the values of $\fx$ and the $\gix$, one is free to study the
behaviour of $W(\al)$ for any set $\ai$. This might be convenient
if one wants to choose between various alternative optimization
prescriptions before actually going to the next iteration. Of course,
if the envisaged new set of `virtual' $\ai$ differs too dramatically from the
\apw\ that were in fact used to obtain the point sample, the error in
the `virtual' $W_i(\al)$ will be large. Since the method outlined in this
paper is quite new, we have not yet studied such embellishments.\\

We shall now discuss the application of the proposed method in practice,
first in a very simple example of stratified-sampling Monte Carlo,
and secondly in an actual event generator.

We start in one dimension,
by integrating the function $f(x)=e^{-x}$ from 0 to $\infty$.
We use three channels, given by
\bqa
g_1(x) & = & \theta(x)\theta(1-x)\;\;,\nl
g_2(x) & = & \theta(x-1)\theta(2-x)\;\;,\nl
g_3(x) & = & \theta(x-2)/(x-1)^2\;\;,
\eqa
in other words, the two uniform distributions between 0 and 1 and
between 1 and 2, respectively, and a `tail' up from 2.
The \apw\ are set to $\alpha_1=\alpha_2=\alpha_3=1/3$ at the
beginning.
We generate 100,000 events to integrate $f(x)$, employing three different
strategies.
In the first case, we integrate without optimization. In the second case,
we optimize once, after the first 1,000 points. In the third case,
we optimize 9 times, after each set of 2,000 points. We monitor the
estimated Monte Carlo error on the integral at every hundredth point.
The results are given in the figure.
\begin{figure}[ht]
\begin{center}
\mbox{\psfig{file=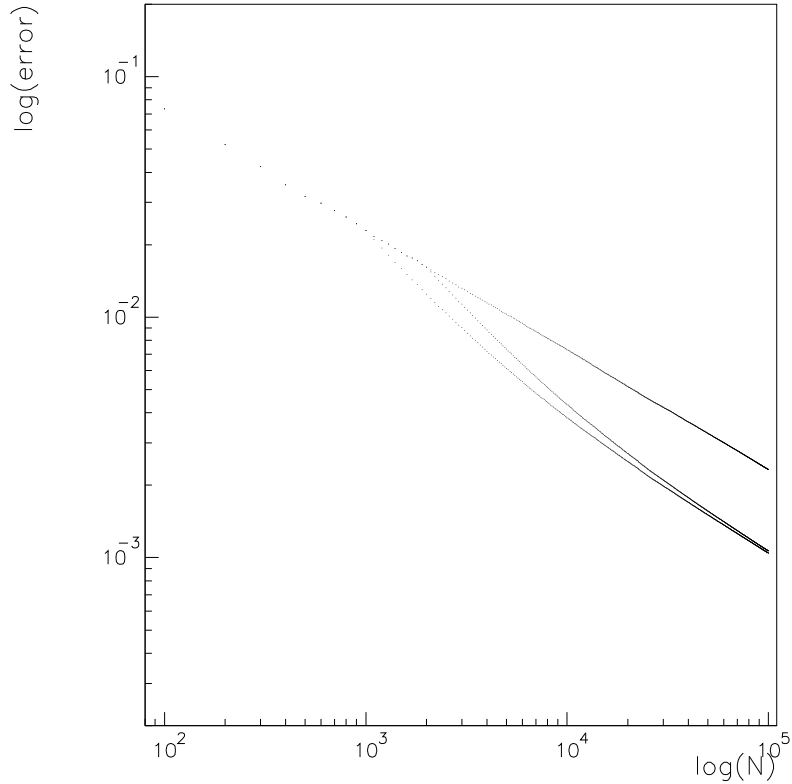,width=12cm,height=12cm}}
\caption[.]{Behaviour of the Monte Carlo error under various optimizations.}
\end{center}
\end{figure}
The case where we do not optimize shows the classical $1/\sqrt{N}$
behaviour (a straight line with slope -1/2 in the double-logarithmic
plot). The effect of optimization in the two other cases is clear:
immediately after optimization, the error decreases faster than before.
Asymptotically, its behaviour is seen to settle again as $1/\sqrt{N}$,
but with a smaller proportionality factor: the variance has been
reduced. The non-linear behaviour just after optimization is due to
the mixture of sets of Monte Carlo points with different variance.
The cases of a single and of multiple optimization steps are seen
to be asymptotically equivalent: both optimizations have brought us close
to the theoretical optimum. In fact, for this simple case we can in fact
compute $\bal_1=0.626$, $\bal_2=0.230$, and $\bal_3=0.144$, whereas
single optimization gave $\al_1=0.623$, $\al_2=0.229$ and $\al_3=0.147$,
and multiple optimization $\al_1=0.636$, $\al_2=0.223$, and $\al_3=0.140$.
In this case, the simpler approach actually happens to be a bit closer to
the theoretical optimum! Since we are dealing here with stratified sampling,
where the approach to the optimum can be very fast ({\it cf.\/} the
discussion above) this is not surprising.

In the above example, the error was reduced by about a factor of two
(equivalent to an increase by a factor 4 in the statistics).
This moderate improvement follows from the fact that our starting
values for the \apw\ happened not to be very far from the optimum
set. We may expect, therefore, that the improvement in the error
may be larger when the $\aib$ are very dissimilar; and this is actually
borne out by our second example, as we shall see.

For every numerical method, simple examples can always be found
where it works very well -- its real merit can only be
judged in a problem of actual interest, such as the following one.
We have applied our optimization approach to the real-life calculation
of a complicated cross section, namely that of electron-positron
annihilation, at LEP200 energies, into an electron-positron
pair and an (electron)neutrino-antineutrino pair. Processes
such as this one
are characterized by a large number of contributing Feynman diagrams,
exhibiting strong peaks in many different regions of phase space,
and the presence of complicated kinematical cuts which render an
analytic treatment impossible. Our Monte Carlo studies of this and similar
processes are reported on in \cite{bkp}, but here we only discuss
the optimization of the \apw\ that we used. For this process
the number of channels $\gix$ is 39,
that initially all have the same a-priori weight
(other four-fermion processes
are described by different numbers of channels, depending on the
kinematics and the appropriate Feynman diagrams). Optimization
is applied after a fixed number of points have been generated, where
the steps between successive optimizations is slowly increased
(this is based on the expected and observed fact that, after an
optimization step, the next improvement needs more statistics to be
effective). At the end of each step, we compute a measure of the
disrepancy between the values of the 39 different $W_i(\al)$:
\bq
D = \max_{i,j}|W_i-W_j|\;\;.
\eq
Hence, $D$ measures how well the set of $\ai$ approximates the behaviour
of the optimal set. At the end of (in this case) 7 steps, we then
determine which of the 7 sets $\ai$ performed the best in terms of
$D$: usually, this is one of the last sets obtained. This set
is then used in the rest of the simulation, with a high-statistics run.
In this way, we feel that we strike a balance between the benefits
of optimization and the possibility that the {\em last\/} set of
$\ai$ obtained is actually a bit worse than a previous one.
The actual numbers come out as follows, for 100,000 Monte Carlo points.
Wihtout optimization of the \apw, the integral (a cross section $\sigma$
of 0.17360 picobarns) has an estimated error $\delta$ of 0.00304 picobarn,
which compares well (that is, to within an order of magnitude)
with the rule of thumb that
$\delta\sim\sigma/\sqrt{N}$: it means that our choice of channels
is reasonably good at describing the various peaks in the cross section.
Now, we turn to the case with optimization. The 7 successive values
of $D$ are given in the table, together with the
value of $N$ at which they were measured.
\begin{center}\begin{tabular}{|c|c|c|}\hline\hline
iter. & $N$ & $D$ \\ \hline
1 & 5,000     & 2.804 \\
2 & 10,000     & 0.416 \\
3 & 15,000     & 0.295 \\
4 & 20,000     & 0.277 \\
5 & 30,000     & 0.254 \\
6 & 40,000     & 0.349 \\
7 & 50,000     & 0.241 \\ \hline\hline \end{tabular}\end{center}
It is seen that $D$ tends to decrease, as expected, except for
iteration 6: in our case we use the set $\ai$ for iteration 7;
had we employed only 6 iterations, we would have chosen the set
for iteration 5. The final part of the integration
consists, then, with a run of 50,000 points,
using the set of $\ai$ obtained in iteration 7.
With this optimization, the integral
comes out as 0.17066 picobarn with an arror of 0.00113 picobarn:
a reduction of nearly a factor of 3. Note that, since the
optimization takes up, a non-negligible fraction of the Monte Carlo points,
the asymptotic improvement is actually somewhat larger (we are not
in the asymptotic regime that can be observed in the figure): the error
would, without optimization, be obtained only for ten times as many Monte
Carlo points. The $\aib$ are actually far from uniform: we started
with $\ai\sim0.025$ for all $i$, and optimization leads to values
ranging from 0.63 down to 0.0000066, a span of 5 orders of
magnitude! This difference between the initial and final set of \apw\
explains the good performance of optimization. Incidentally, note that,
in the Monte Carlo program, points sometimes have to be assigned
zero weight (for instance, if they fall outside the experimentally defined
cuts): without optimization, we have 40,848 such events out of 100,000:
with optimization, there are 29,111 such points left -- another indication
of self-adjustement at work.\\

In conclusion, we have described a simple method for automatically
improving the performance of multichannel Monte Carlo, that is,
any Monte Carlo integration where a decision
is made at some point, using predetermined probabilities,
on where, or how, to choose the next point
(stratified and importance sampling, respectively). The
strategy is very modest in terms of time or memory requirements,
and has been seen to perform well in a realistic application to
a very nontrivial physical problem. We feel that its use deserves
consideration in complicated event-generating simulation programs
such as are used in high-energy physics.

\end{document}